\documentclass[a4paper,11pt]{article}

\usepackage{contribution}
\usepackage{dsfont}
\usepackage{epsfig}

\newcommand{\weblink}[2][]{%
    \ifthenelse{\equal{#1}{}}%
    {\textnormal{\url{#2}}}%
    {\textnormal{\href{#2}{#1}}}%
}

\def\beq{\begin{equation}}
\def\eeq#1{\label{#1}\end{equation}}
\def\eeqn{\end{equation}}

\def\beqa{\begin{eqnarray}}
\def\eeqa#1{\label{#1}\end{eqnarray}}
\def\eeqan{\end{eqnarray}}

\let\bar=\overbar

\def\Dslash{\not{\hbox{\kern-4pt $D$}}}
\def\dslash{\not{\hbox{\kern-2pt $\del$}}}

\def\msb{{\bar{\ssstyle M \kern -1pt S}}}

\newcommand{\contribution}[7][]{%
  \clearpage
  \thispagestyle{plain}
  \ifthenelse{\equal{#1}{}}
  {\hypersetup{pdftitle={#2}}}
  {\hypersetup{pdftitle={#1}}}
  \hypersetup{pdfauthor={{#3} {#4}}}
  {\centering\normalfont\LARGE\bfseries\sffamily #2 \par\nobreak}
  \lhead{}
  \chead{%
    \textit{\footnotesize XIV International Conference on Hadron Spectroscopy
      (\weblink[\textit{hadron2011}]{http://www.hadron2011.de}), 13-17 June 2011, Munich, Germany}%
  }
  \rhead{}
  \bigskip
  \begin{center}
    {#3} {#4}\ifthenelse{\equal{#6}{}}{}{\footnote{\weblink[#6]{mailto:#6}}}
    \ifthenelse{\equal{#7}{}}{}{#7} \\
    \textit{#5}
  \end{center}
  \bigskip
}

\renewcommand{\abstract}[1]{%
  \begin{center}
    \begin{minipage}{0.85\textwidth}
      \begin{footnotesize}
        #1
      \end{footnotesize}
    \end{minipage}
  \end{center}
  \bigskip
}

\begin{document}


%
%
%
%

{  


%
\contribution[scalar mesons in a finite volume]
{Chiral unitary theory of scalar mesons in a finite volume}
{E.}{Oset}  
{$^1$Departamento de F\'isica Te\'orica and IFIC, Centro Mixto\\
Universidad de
Valencia-CSIC, Institutos de Investigaci\'on de Paterna, Aptdo. 22085,
46071
Valencia, Spain\\
$^2$Helmholtz-Institut f\"ur Strahlen- und Kernphysik (Theorie) and
Bethe Center
for Theoretical Physics,  Universit\"at Bonn, Nu\ss allee 14-16, D-53115
Bonn, Germany\\
$^3$JCHP, IAS-4, IKP-3, Forschungszentrum J\"ulich, D-52425 J\"ulich,
Germany
} 
{}
{$^1$, M.~D\"oring$^2$, U.-G. ~Mei\ss ner$^{2,3}$ and A. Rusetsky$^2$}




\abstract{We develop a scheme for the extraction of the properties of the scalar mesons
$f_0(600)$, $f_0(980)$, and $a_0(980)$ from lattice QCD data.  This scheme is based
on a two-channel  chiral unitary approach with fully relativistic propagators
in a finite volume. In order to discuss the feasibility of finding the mass and
width of the scalar resonances, we analyze synthetic lattice data with a fixed
error assigned, and show that the framework can be indeed used for an accurate
determination of resonance pole positions in the multi-channel scattering.
}
%

\section{Introduction}

We present a method to calculate energy levels for $\pi \pi$ and $K \bar{K}$ in a finite volume using for this purpose the chiral unitary approach that produces the $f_0(600)$, $f_0(980)$, and $a_0(980)$ resonances as dynamically generated in the continuum. After this we face the inverse problem: assuming that the levels obtained in the finite box are lattice data we determine the phase shifts in the continuum and extract the resonance properties \cite{misha}.

\section{The formalism}

In the chiral unitary approach we obtain the T-matrix for the $\pi \pi $ and $K \bar{K}$ coupled channels by means of the Bethe Salpeter equation
\begin{equation}
T=[1-VG]^{-1}V
\label{bse}
\end{equation}
where V is the 2x2 matrix of the transition potential and G, a diagonal matrix is the loop function of the two meson propagators
\begin{eqnarray}
G_j&=&\int\limits^{|\vec q|<q_{\rm max}}
\frac{d^3\vec q}{(2\pi)^3}\frac{1}{2\omega_1(\vec q)\,\omega_2(\vec q)}
\frac{\omega_1(\vec q)+\omega_2(\vec q)}
{E^2-(\omega_1(\vec q)+\omega_2(\vec q))^2+i\epsilon},
\nonumber \\[2mm] 
\omega_{1,2}(\vec q)&=&\sqrt{m_{1,2}^2+\vec q^2}\, .
\label{prop_cont}
\end{eqnarray}

In a box of length L the energy levels are given by the poles of the T matrix of eq. (\ref{bse}) substituting $G$ by $\tilde G$ given by 
\begin{eqnarray}
\tilde G_{j}&=&\frac{1}{L^3}\sum_{\vec q}^{|\vec q|<q_{\rm max}}
\frac{1}{2\omega_1(\vec q)\,\omega_2(\vec q)}\,\,
\frac{\omega_1(\vec q)+\omega_2(\vec q)}
{E^2-(\omega_1(\vec q)+\omega_2(\vec q))^2},
\nonumber \\[2mm] 
\vec q&=&\frac{2\pi}{L}\,\vec n,
\quad\vec n\in \mathds{Z}^3 \ .
\label{tildeg}
\end{eqnarray}

As we can see, all we have done is to replace the integral by a discrete sum over the free eigenvalues of the box given by the periodic boundary conditions.

In one channel, the poles of $T$ in the box are obtained when 
\begin{equation}
V^{-1}(E)-\tilde G(E)=0\, .
\label{vmin1}
\end{equation} 

Then the $T$ matrix in the continuum for the energies eigenvalues of the box can be obtained by means of 
\begin{equation}
T(E)=\left(V^{-1}(E)-G(E)\right)^{-1}= \left(\tilde G(E)-G(E)\right)^{-1} \ . 
\label{extracted_1_channel}
\end{equation}

By changing the value of L one can achieve that different energies appear as eigenvalues of the box and then obtain the $T$ matrix in the continuum for any desired energy, via eq. (\ref{extracted_1_channel}). As shown in \cite{misha} this equation is nothing else than L\"uscher's formula \cite{luscher}. However, for smaller values of $L$, differences emerge which are
due to the presence of the relativistic propagators in the loops
(see Ref. \cite{misha} for a detailed discussion on this issue). 

The main purpose of \cite{misha} was to extend the idea of L\"uscher to two channels, following the lines of \cite{akaki}. For this purpose we propose several methods in \cite{misha}, but we will outline only one here, which is practical and at reach by present lattice calculations. 

Let us assume that we have the lattice data of fig. 1, which have been obtained from the chiral unitary approach in the finite box, and we want to obtain the $\pi \pi$ and $K \bar{K}$ phase shifts in the continuum from these data. For this purpose we take the levels 2 and 3 of Fig. 1 and a few energies from them, associating an error of 10 MeV to these energies. Then we make a fit to these data assuming that we have a potential, suggested by the chiral unitary approach, of the type
\begin{equation}
V_{ij}=a_{ij}+b_{ij}(s-4M_K^2)\, .
\label{fitv}
\end{equation}

The levels in the box with two channels are obtained from the poles of the $T$ matrix which come from the condition that the determinant of ($1-VG$) is zero,
\begin{equation}
\label{eq:det}
\det(\mathds{1}-V\tilde G)=1-V_{11}\tilde G_1-V_{22}\tilde G_2
+(V_{11}V_{22}-V_{12}^2)\tilde G_1\tilde G_2=0\, .
\end{equation}

\begin{figure}
\begin{center}
\includegraphics[width=0.38\textwidth]{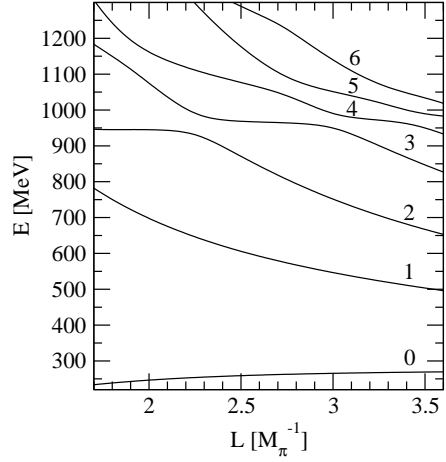}
\caption{Energy levels as functions of the cubic box size $L$, derived 
from the chiral unitary approach of Ref.~\cite{npa} and using 
$\,\tilde G$ from Eq.~(\ref{tildeg}).}
\label{fig:levels_nonsmooth}
\end{center}
\end{figure}

\begin{figure}
\begin{center}
\includegraphics[width=0.48\textwidth]{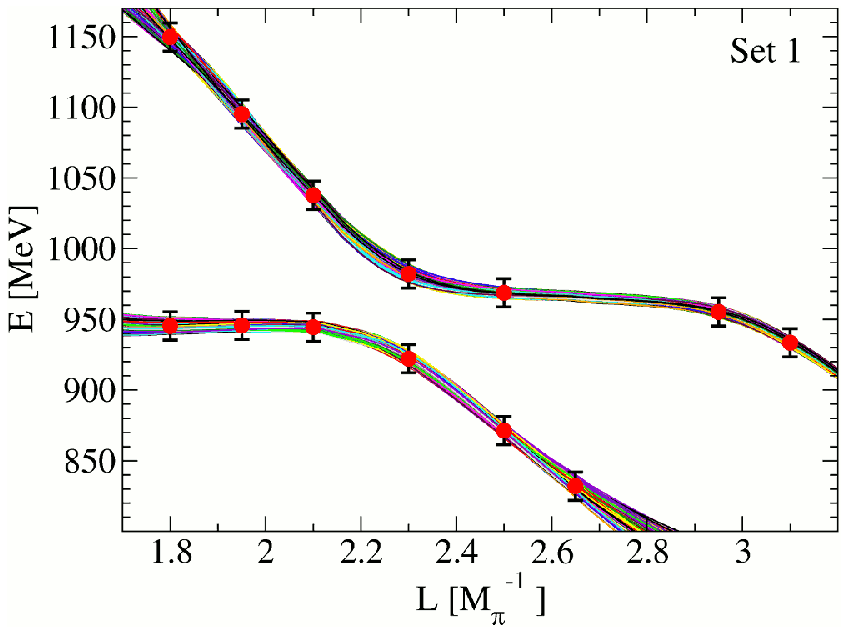}\hspace{0.5cm}
\includegraphics[width=0.45\textwidth]{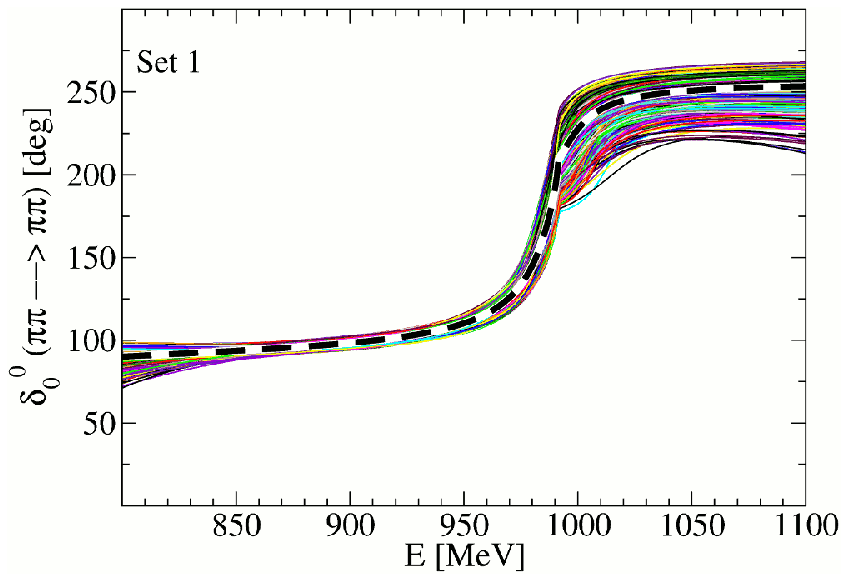}\\
\caption{Left: Generated data points (10 MeV error, 13 points) with periodic b.c.;    Fits that fulfill the $\chi_{\rm
best}^2+1$ criterion are also shown (bands). Right: Extracted phase shifts corresponding to the set shown in the figure to the left.
  Dashed line: The calculated phase shifts by using the
approach of  Ref.~\cite{npa}; Bands: Reconstruction of these phase shift.}
\label{fig:errors2}
\end{center}
\end{figure}

As one can see in fig. \ref{fig:errors2}, the method shows that one can reconstruct the phase shifts in the continuum with an acceptable accuracy and from there the $f_0(980)$ resonance. The same can be done for the $a_0(980)$ and the $f_0(600)$. As lessons that one draws from the work we can say that the two channel method is necessary for the description of the $f_0(980)$ and $a_0(980)$ resonances, the analysis with only the $\pi \pi$ channel leading to inaccurate results and incorrect conclusions. Another lesson learned is that the flattening of a level in the box as a funtion of L  is not a guarantee that this energy corresponds to a resonance. The flattening can occur around the threshold of a new channel without it corresponding to a resonance in the continuum. The method developed in \cite{misha} should be very helpful in the analysis of future lattice data for groups looking for hadron spectra from lattice QCD \cite{lattice}.  

%




}  



\begin{thebibliography}{99}

\bibitem{misha}
 M.~D\"oring, U.~-G.~Mei{\ss}ner, E.~Oset, A.~Rusetsky,
 [arXiv:1107.3988 [hep-lat]].
 
\bibitem{luscher}
 M.~L\"uscher,
 Commun.\ Math.\ Phys.\  {\bf 105} (1986) 153 (1986).

\bibitem{akaki}
 V.~Bernard, M.~Lage, U.-G.~Mei{\ss}ner and A.~Rusetsky,
 JHEP {\bf 1101} (2011) 019.
 
\bibitem{npa}
  J.~A.~Oller, E.~Oset,
  Nucl.\ Phys.\  {\bf A620}, 438-456 (1997).
  

 
\bibitem{lattice}
 Y.~Nakahara, M.~Asakawa, T.~Hatsuda,
 Phys.\ Rev.\  {\bf D60} (1999)  091503;
 K.~Sasaki, S.~Sasaki and T.~Hatsuda,
 Phys.\ Lett.\  B {\bf 623} (2005) 208;
 N.~Mathur, A.~Alexandru, Y.~Chen {\it et al.},
 Phys.\ Rev.\  {\bf D76} (2007) 114505;
 S.~Basak, R.~G.~Edwards, G.~T.~Fleming {\it et al.},
 Phys.\ Rev.\  {\bf D76} (2007) 074504; 
 J.~Bulava, R.~G.~Edwards, E.~Engelson {\it et al.},
 Phys.\ Rev.\  {\bf D82} (2010) 014507;
 C.~Morningstar, A.~Bell, J.~Bulava {\it et al.},
 AIP Conf.\ Proc.\  {\bf 1257} (2010) 779;
 J.~Foley, J.~Bulava, K.~J.~Juge {\it et al.},
 AIP Conf.\ Proc.\  {\bf 1257} (2010) 789;
 M.~G.~Alford and R.~L.~Jaffe,
 Nucl.\ Phys.\  B {\bf 578} (2000) 367;
 T.~Kunihiro, S.~Muroya, A.~Nakamura, C.~Nonaka, M.~Sekiguchi and H.~Wada
                 [SCALAR Collaboration],
 Phys.\ Rev.\  D {\bf 70} (2004) 034504;
H.~Suganuma, K.~Tsumura, N.~Ishii and F.~Okiharu,
 PoS {\bf LAT2005} (2006) 070;
 Prog.\ Theor.\ Phys.\ Suppl.\  {\bf 168} (2007) 168; 
 C.~McNeile and C.~Michael  [UKQCD Collaboration],
 Phys.\ Rev.\  D {\bf 74} (2006) 014508;
A.~Hart, C.~McNeile, C.~Michael and J.~Pickavance  [UKQCD Collaboration],
 Phys.\ Rev.\  D {\bf 74} (2006) 114504;
 H.~Wada, T.~Kunihiro, S.~Muroya, A.~Nakamura, C.~Nonaka and M.~Sekiguchi,
 Phys.\ Lett.\  B {\bf 652} (2007) 250;
S.~Prelovsek, C.~Dawson, T.~Izubuchi, K.~Orginos and A.~Soni,
 Phys.\ Rev.\  D {\bf 70} (2004) 094503;
S.~Prelovsek, T.~Draper, C.~B.~Lang, M.~Limmer, K.~F.~Liu, N.~Mathur 
 and D.~Mohler,
 arXiv:1002.0193 [hep-ph].
 



\end{thebibliography}
\end{document}